\documentclass[a4paper, amsfonts, amssymb, amsmath, twoside, twocolumn, showkeys, nofootinbib, aps, pra, 10pt]{revtex4-2}
\usepackage[english]{babel}
\usepackage[utf8]{inputenc}
\usepackage[colorinlistoftodos, color=green!40, prependcaption]{todonotes}
\usepackage{amsthm}
\usepackage{mathtools}
\usepackage{dsfont}
\usepackage{physics}
\usepackage{xcolor}
\usepackage{graphicx}
\usepackage[paperwidth=210mm, paperheight=297mm, centering, hmargin=2cm, vmargin=2.5cm]{geometry}
\usepackage{adjustbox}
\usepackage{placeins}
\usepackage[T1]{fontenc}
\usepackage{lipsum}
\usepackage{csquotes}
\usepackage{soul} 
\setstcolor{red}
\usepackage[pdftex, pdftitle={Article}, pdfauthor={Author}]{hyperref} 
\newcommand{\PauliScatt}[4]{\lambda \begin{pmatrix} #1 & #2 \\ #3 & #4 \end{pmatrix}}
\newcommand{\IntScatt}[4]{\xi \begin{pmatrix} #1 & #2 \\ #3 & #4 \end{pmatrix}}

\begin{document}
\title{Composite Bosons Superposition Ansatz Approach to One-Dimensional Trapped Few-Fermion Systems}

\author{Francisco Figueiredo}
    \email[Correspondence email address: ]{frfig@cbpf.br}
    \affiliation{Centro Brasileiro de Pesquisas Físicas, Rua Dr. Xavier Sigaud 150, 22290-180 Rio de Janeiro, Brazil}

\author{Itzhak Roditi}
    \email[Correspondence email address: ]{roditi@cbpf.br}
    \affiliation{Centro Brasileiro de Pesquisas Físicas, Rua Dr. Xavier Sigaud 150, 22290-180 Rio de Janeiro, Brazil}

\date{\today} 

\begin{abstract}
    Ultra-cold atomic systems provide a versatile platform for exploring quantum phenomena, offering tunable interactions and diverse trapping geometries. In this study, we investigate a one-dimensional system of trapped fermionic atoms using the composite boson formalism, which describes pairs of opposite-spin fermions as cobosons (short for composite bosons). {We solve the Schrödinger equation for a $2\uparrow + 2\downarrow$ fermionic system across a broad range of attractive and repulsive interactions by constructing the system’s wave function as a superposition of two-coboson states.} We determine key observables such as particle density profiles and two-body correlations. {The density profiles calculated agree with both the Tonks-Girardeau and Lieb-Lineger limits and provide a portrait of the transition between those regimes. In the strong repulsive regime, the ansatz captures the Fridel-Wigner transition characterized by a doubling of the peaks in the density profiles.} Additionally, we compute the low-lying energy spectrum and estimate the pairing gap. {Our results highlight the usefulness of the coboson superposition ansatz for exploring quantum phenomena in strongly correlated few-body systems.}
\end{abstract}

\keywords{Cold-atoms, few-fermion systems, composite bosons}

\maketitle

\section{Introduction} \label{sec:introduction}

Ultra-cold atomic systems have proven to be a powerful platform for studying fundamental quantum phenomena, as they can be experimentally realized in a wide variety of trapping geometries with tunable interactions and diverse particle types, covering regimes ranging from few to many body \cite{bloch2008many, Chin2010}. Interactions can be finely tuned across a wide range via magnetic Feshbach resonances \cite{Chin2010}. Confining geometries come in various forms, such as three-dimensional geometries \cite{Bourdel2004}, quasi two-dimensional arrangements \cite{BEC-BCS2D_Jochim2015}, quasi one-dimensional configurations \cite{zurn2012fermionization}, double-well structures \cite{murmann2015two} and lattices \cite{lewenstein2012ultracold}, to name a few. Two seminal milestones in the field are the experimental achievements of the condensation of bosonic atoms \cite{BEC_1995} and of fermionic molecules \cite{Zwierlein2003, Jochim2101}. Furthermore, quantum simulators \cite{bloch2012quantum, gross2017quantum} and quantum metrology \cite{pezze2018quantum, szigeti2021improving} can be referred to as important areas of quantum technology where such systems are gaining increasing notoriety.

There is a growing body of research, both experimental and theoretical, focused on systems with quasi-one-dimensional trapping geometries \cite{volosniev2014strongly, guan2013fermi, sowinski2019one, minguzzi2022strongly, mistakidis2023few}. Fermionic systems in such geometries are believed to exhibit unconventional pairing phases \cite{sowinski2021few, pkecak2020signatures, lydzba2020unconventional, dobrzyniecki2021unconventional, pkecak2022unconventional}, since reduced dimensionality can induce strong correlations between the constituent particles. Such systems open the possibility for studying the transition from few- to many-body regimes \cite{serwane2011deterministic, wenz2013few, grining2015crossover, rammelmuller2017evolution} and can also be employed to study the emergence of quantum phase transitions \cite{rammelmuller2023magnetic}. 

A variety of theoretical approaches have been applied to study such strongly correlated systems, including methods from quantum chemistry such as exact diagonalization \cite{rontani2006full, rojo2020static, sowinski2021few, pkecak2016few, pkecak2022unconventional, rojo2022direct, rammelmuller2023modular} and coupled cluster theories \cite{grining2015crossover,grining2015many}, as well as Bethe ansatz \cite{rubeni2012two, guan2013fermi}, variational ansatz \cite{wilson2014geometric, kudla2015pairing, patton2017trapped, koscik2018variational}, Multi Configuration Time Dependent Hartree \cite{brouzos2013two}, Quantum Monte Carlo \cite{casula2008quantum, rammelmuller2017evolution}, Density Matrix Renormalization Group \cite{bellotti2017comparing}, among others. 

Here, we develop the study of a one-dimensional trapped system of fermionic atoms through an approach based on the formalism of composite bosons (cobosons, for short) \cite{COMBESCOT2008, CombescotExcitonsCooperPairs}. Cobosons are states of pairs of opposite-spin fermions. The coboson formalism arises in the early 2000's from the intention of understanding the approximated bosonic behavior of many-exciton systems \cite{Combescot2001, Combescot2003} and also to present a theoretical description of such composite particles that fully takes into account the fermionic nature of its components, as opposed to approaches involving bosonization procedures \cite{combescot2002effective, combescot2005many}. Furthermore, the formalism can be extended to investigate arbitrary composite particles \cite{combescot2010general} and settings with finite temperature \cite{CobosonZ_Shiau2015, combescot2011general_finite_T}.

A simple wave function ansatz for $N$ pairs of fermions, which shall be referred to as the standard coboson ansatz, is one in which all pairs occupy the ground state. This ansatz has been used across a range of many different applications. In semiconductor physics, this ansatz provides a valid approximation of the ground-state of $N$ excitons in the low density limit \cite{Combescot2001, Combescot2003, keldysh1968collective}. Going beyond excitonic systems with long-range interactions, the latter was also applied to settings with short-range interactions like BCS superconductivity \cite{combescot2011Richardson, zhu2012bcs, combescot2013bcs} - where the BCS wave function has the same form as the ansatz when restricted to the $N$-pair subspace \cite{combescot2013bcs} - and to ultra-cold atomic systems \cite{Combescot2016, Shiau2016, Bouvrie2017, Bouvrie-Cuestas-Roditi-Majtey_2019}.

Going past the standard ansatz, the formalism allows one to construct more complex wave functions. The idea is use a basis of single-pair states (cobosons) to build a basis for the $N$-pair subspace and to solve the Schrödinger equation with respect to this space. The wave functions of the system will be given as a superposition of $N$-coboson states. Such wave function ansatz has been applied, for example, to determine the electronic structure and absorption spectrum of biexcitons \cite{shiau2013electronic} and also to calculations of dimer-dimer and atom-dimer scattering lengths \cite{Shiau2016, shiau2019coboson}. More recently, there are applications to systems of strongly bound fermion pairs in one dimension \cite{cuestas2022strongly, jimenez2023composite}. Approaches based on pair wave functions have also {been gaining} attention as alternatives to standard single-particle electronic structure methods \cite{tecmer2022geminal}.

In this work, we consider a one-dimensional harmonically trapped system comprising two pairs of opposite spin-$1/2$ fermions interacting via a contact potential. This model has a well known analytic solution for a single pair \cite{avakian1987spectroscopy, busch1998two, franke2003coherence}. We determine a numerically exact representation of the single-pair eigenstates of the system with respect to a harmonic oscillator basis. With such representation, we employ a many-coboson superposition ansatz to solve the Schrödinger equation for both, attractive and repulsive, regimes of interaction. 
In order to characterize the ground-state of the system, we analyze key one-body and two-body observables like density profiles and two-body correlators \cite{sowinski2021few}. The density profiles calculated from this ansatz agree with both the Tonks-Girardeau and Lieb-Lineger limits \cite{grining2015crossover,grining2015many} and provide a portrait of the transition between those regimes.

We also determine the low-lying energy spectrum of the system for weak interactions and estimate the few-body analogue of the pairing gap \cite{d2015pairing}, which is a critical quantity in many-body physics. In few-body systems, this energy gap can serve as a key indicator of the formation of bound states (such as Cooper pairs) and provides insights into the transition between weakly and strongly interacting regimes. By solving for the low-lying states, it is possible to estimate the pairing gap through the spin gap \cite{rammelmuller2023magnetic} and analyze its evolution as the interaction strength is varied. We consider this approach a sensible method for tackling strongly correlated few-body systems, as it yields reasonable results with a relatively modest number of terms in the superposition.

The paper is organized as follows. In section \ref{sec:model}, we introduce the system Hamiltonian and its main symmetries. Then, in section \ref{sec:formalism}, we provide a brief outline of the composite boson formalism evidencing the key mathematical objects. Section \ref{sec:apply} contains the description of the construction of the coboson superposition ansatz for a balanced $2\uparrow + 2\downarrow$ fermion system. In section \ref{sec:results}, we give an analysis of key one-body and two-body observables of the system across a range of interactions. Section \ref{sec:conclusion} concludes the work, summarizing the key aspects and providing future perspectives. After the conclusion, we provide formulas for key quantities and observables in appendix \ref{sec:appendix A}.

\section{The model} \label{sec:model}

We will work with a one-dimensional trapped fermionic systems of particles belonging to one of two spin-$\frac{1}{2}$ species. The system is modeled by the following many-body Hamiltonian
\begin{align}\label{Many-body-Hamiltonian}
    \hat{\mathcal{H}} = \sum\limits_{\sigma} \int dz\textrm{ }\hat{\Psi}^\dagger_{\sigma}(z)H_0\hat{\Psi}_{\sigma}(z)  + \hat{V},
\end{align}
where $\sigma \in \{\uparrow, \downarrow\}$ labels the two fermionic species and $\hat{\Psi}^\dagger_{\sigma}(z)$ is the field operator that creates a sharply localized state in the position-spin representation, i.e. $\hat{\Psi}^\dagger_{\sigma}(z)\ket{0} = \ket{z, \sigma}$ with $\ket{0}$ being the vacuum state. The free part of the Hamiltonian is given by
\begin{align}
    H_0 = -\frac{1}{2}\frac{\partial^2}{\partial z^2} + \frac{1}{2} z^2.
\end{align}
Throughout the paper, lengths and energies are expressed in units of $a_0 = \sqrt{\frac{\hbar}{m\omega}}$ and $\hbar\omega$, respectively, with $m$ being the mass of the fermions and $\omega$ the characteristic frequency of the harmonic trap confining the particles. The different fermionic species are considered to have the same mass.

The fermions interact through a short-range $\delta$-potential 
\begin{align}\label{Many-body-V_int}
    \hat{V} 
    = g\int dz\textrm{ }
    \hat{\Psi}^\dagger_{\uparrow}(z)\hat{\Psi}^\dagger_{\downarrow}(z)
    \hat{\Psi}_{\downarrow}(z)\hat{\Psi}_{\uparrow}(z)
\end{align}
with dimensionless coupling constant $g$ taking values ranging from $-\infty$ to $+\infty$ representing, respectively, the attractive and repulsive regimes. This one-dimensional model can describe a system in a quasi-1D cigar-shaped trap \cite{zurn2012fermionization}. The $\delta$-potential is widely used to model the s-wave interaction regime of ultra-cold atoms \cite{olshanii1998atomic, idziaszek2006analytical}.

The Hamiltonian (\ref{Many-body-Hamiltonian}) with interaction (\ref{Many-body-V_int}) has some important  symmetries. First, the Hamiltonian commutes with the operators $\hat{N}_\sigma = \int dz\textrm{ }\hat{\Psi}^\dagger_{\sigma}(z)\hat{\Psi}_{\sigma}(z)$ which implies the conservation of particle number in each component. Since the confining potential is quadratic and spin-independent, the center-of-mass motion can be decoupled from the internal motion, which constitutes a $U(1)$ symmetry. Finally, the Hamiltonian is invariant under spatial inversion $\mathcal{P}$ and also under a spin-flip $U_\mathcal{T}$ respectively being defined by
\begin{align}
    &\mathcal{P}\hat{\Psi}^\dagger_{\sigma}(z)\mathcal{P}^{-1} = \hat{\Psi}^\dagger_{\sigma}(-z)\label{Parity_Psi} \\
    &U_\mathcal{T}\hat{\Psi}^\dagger_{\uparrow\downarrow}(z){U_\mathcal{T}}^{-1} = \hat{\Psi}^\dagger_{\downarrow\uparrow}(z).\label{SpinFlip_Psi}
\end{align}
We denote the spin-flip operator by $U_\mathcal{T}$ since it is an unitary operator entering the time reversal operator $\mathcal{T}$ for spin-$\frac{1}{2}$ particles, see for example Ref. \cite{dresselhaus2007group}.

\section{Outline of the coboson formalism} \label{sec:formalism}

In this section, we provide an brief outline of the composite boson formalism. The exposition is mainly based on \cite{Combescot2016, Shiau2016, CombescotExcitonsCooperPairs}.

The fundamental building blocks of the formalism are the states of fermion pairs, i.e. coboson states. A generic state of an $\uparrow\downarrow$-pair can be represented as
\begin{align}\label{generic_pair_state}
\ket{\Phi} = \int dz_\alpha dz_\beta \textrm{ } \varPhi(z_\alpha, z_\beta) \hat{\Psi}^\dagger_{\uparrow}(z_\alpha)\hat{\Psi}^\dagger_{\downarrow}(z_\beta)\ket{0},
\end{align}
where $\varPhi(z_\alpha,z_\beta)$ is the spatial wave function of the pair in the Schrödinger representation. We employ the indexes $\alpha, \beta$ to label the $\uparrow$ and $\downarrow$ spin species, respectively. One can define the creation and annihilation operators for this state $\ket{\Phi}$ as
\begin{align}\label{coboson_spin-spatial}
    B^\dagger_{\ket{\Phi}} &:= \int dz_\alpha dz_\beta \textrm{ } \varPhi(z_\alpha, z_\beta) \hat{\Psi}^\dagger_{\uparrow}(z_\alpha)\hat{\Psi}^\dagger_{\downarrow}(z_\beta)\\
    B_{\ket{\Phi}} &:= \left(B^\dagger_{\ket{\Phi}}\right)^\dagger.
\end{align} 

The second quantization 'recipe' to construct states with many of these pairs involves applying these operators successively onto the vacuum. The rules for how these successive applications 'stack up' the states (quantum statistics) are contained in commutation/anti-commutation an algebra of such operators. The dynamical aspects are contained in the commutation relations with the system's Hamiltonian $\hat{\mathcal{H}}$ \cite{Combescot2016}.

One possible choice to describe the $\uparrow\downarrow$-pair subspace of the system is to use a basis of eigenstates of the free part of (\ref{Many-body-Hamiltonian}). Let $\left\{c^\dagger_{n,\sigma}\ket{0}\right\}$ be a basis of free fermion eigenstates, i.e. simple harmonic oscillator states with orbital wave function given by 
\begin{align}\label{HO_func_n}
\varphi_{n}(z) = (\pi^{1/2} 2^n n!)^{-1/2} H_{n}(z) e^{-z^2/2}  
\end{align}
where $H_{n}(z)$ is a Hermite polynomial and $n \in \mathbb{N}_0$. From this single-particle basis, one can build a basis $\{ d^\dagger_{n_{\alpha}n_{\beta}}\ket{0}\}$ for the $\uparrow\downarrow$-pair subspace, where  $d^\dagger_{n_{\alpha}n_{\beta}}:= c^\dagger_{n_\alpha,\uparrow}c^\dagger_{n_\beta,\downarrow}$ and 
\begin{align}\label{H0_pair_eigen_eq}
	\left( \hat{\mathcal{H}}_0 -\varepsilon_{n_\alpha n_\beta} \right)d^\dagger_{n_{\alpha} n_{\beta}}\ket{0} = 0,
\end{align}
with $\varepsilon_{n_\alpha n_\beta} = n_\alpha + n_\beta + 1$. As pointed out, the rules of 'pilling up' the operators $d^\dagger_{n_{\alpha} n_{\beta}}$ in order to build many-pair states are contained in the commutation algebra of such operators. We would like, however, to develop the formalism with respect to a special basis of correlated pairs, the single-pair eigenstates of the system.

The set of single-pair eigenstates of the system,
\begin{equation}\label{cobosons_eq_2quantz}
	\left(\hat{\mathcal{H}} - E_i\right)\ket{\Phi_i} = 0,
\end{equation}
with $E_i$ being the $i$-th eigenstate energy, also constitutes a basis for the $\uparrow\downarrow$-pair subspace. One then defines the $i$-th coboson eigenstate creation operator as $B^\dagger_i := B^\dagger_{\ket{\Phi_i}}$, following equation (\ref{coboson_spin-spatial}).

The commutation relations characterizing the kinematics of many-coboson states are 
\begin{align}
    &[B_m,B_i^\dagger] = \delta_{mi} - \mathcal{D}_{mi}\label{commut_Kin_1}\\
    &[B_m,B_i] = 0 = [ B_m^\dagger,B_i^\dagger]\label{commut_Kin_2}\\
    &[\mathcal{D}_{mi},B_j^\dagger] = \sum\limits_n \left[ \lambda\begin{pmatrix} n & j \\ m & i \end{pmatrix} + \lambda\begin{pmatrix} n & i \\ m & j \end{pmatrix}  \right]B_n^\dagger \label{commut_Kin_3}
\end{align}
Relations (\ref{commut_Kin_1})-(\ref{commut_Kin_2}) are almost identical to those of ideal bosons with the difference being found in the term including the operator $\mathcal{D}_{mi}$. This operator quantifies the deviation from ideal bosonic behavior. Its explicit form is not of direct use for this work, however we only highlight the key property that it annihilates the vacuum, i.e. $\mathcal{D}_{mi}\ket{0} = 0$. The commutation relation (\ref{commut_Kin_3}) bears the so-called Pauli scattering
\begin{align}\label{PauliScatt_eq}
    \lambda\begin{pmatrix} n & j \\ m & i \end{pmatrix} 
=\int d^4z\textrm{ } 
&{\varPhi_m(z_{\alpha_1},z_{\beta_2})}^*
{\varPhi_n(z_{\alpha_2},z_{\beta_1})}^* \nonumber \\
&\times\varPhi_i(z_{\alpha_1},z_{\beta_1})
\varPhi_j(z_{\alpha_2},z_{\beta_2}), 
\end{align}
which quantifies the contribution to the overlap of pair states due to the exchange of fermions $\beta_1 \leftrightarrow \beta_2$. The notation $\lambda\begin{pmatrix} n & j \\ m & i \end{pmatrix}$ is intended to suggest that this quantity is related to the scattering of two cobosons, which start in states $i, j$ and end in states $m,n$.

A important kinematic result is that the $N>1$ pair subspace is spanned by the set of many-coboson states $\{B_{i_1}^\dagger\dots B_{i_N}^\dagger\ket{0}\}$, however they form an overcomplete set satisfying
\begin{equation}\label{cob_identity}
    \mathds{1}_N = \frac{1}{(N!)^2}\sum\limits_{i_1,\dots,i_N}B_{i_1}^\dagger\dots B_{i_N}^\dagger\ket{0}\bra{0}B_{i_N}\dots B_{i_1},
\end{equation}
with $\mathds{1}_N$ being the $N$-pair subspace identity \cite{COMBESCOT2008}.

The dynamics within the formalism is characterized by 
\begin{align}
    &[\hat{\mathcal{H}},B_i^\dagger] = E_iB_i^\dagger + \mathcal{V}_i^\dagger \label{H_B_commut}\\
    &[\mathcal{V}_i^\dagger,B_j^\dagger] = \sum\limits_{m,n}\xi\begin{pmatrix} n & j \\ m & i \end{pmatrix}B_m^\dagger B_n^\dagger\label{V_B_commut}
\end{align} 
where $E_i$ is the eigenenergy of ${B_i^\dagger\ket{0}}$. Relations (\ref{H_B_commut})-(\ref{V_B_commut}) define two other key quantities of the formalism, the creation potential $\mathcal{V}_i^\dagger$ and the interaction scattering $\xi$. For systems where only fermions from different spin species interact, which is the case of our interaction (\ref{Many-body-V_int}), it can be shown that
\begin{align}\label{Creation_potential_cold-atoms}
	\mathcal{V}^\dagger_i = -\sum\limits_{m,m'}B_{m}^\dagger V_{m m'} \mathcal{D}_{m'i},
\end{align}
where $V_{m m'} = \bra{\Phi_{m}}\hat{V}\ket{\Phi_{m'}}$ are the matrix elements of the potential with respect to the single-pair eigenstate basis \cite{Combescot2016}. This object also has the property that it annihilates the vacuum, i.e. $\mathcal{V}_i^\dagger\ket{0} = 0$.
It follows from equation (\ref{Creation_potential_cold-atoms}) that the interaction scattering assumes the form
\begin{align}\label{interaction_Scatt}
    \xi\begin{pmatrix} n & j \\ m & i \end{pmatrix} &= 
    -\frac{1}{2}\left[\sum\limits_{m'}V_{m m'} \lambda\begin{pmatrix} n & j \\ m' & i \end{pmatrix} + \right. \nonumber\\
    &+\left.\sum\limits_{n'}V_{n n'} \lambda\begin{pmatrix} n' & j \\ m & i \end{pmatrix} + (i\leftrightarrow j)
    \right ].
\end{align}
This object quantifies how the direct interaction between the constituent fermions of different cobosons influences their scattering. The Pauli and interaction scatterings depend only on the eigenstates of a single pair and the interaction potential. We also introduce the following combination of both scatterings
\begin{align}\label{in_interaction_Scatt}
    \xi^{in}\begin{pmatrix} n & j \\ m & i \end{pmatrix} := \sum\limits_{r,s} \lambda\begin{pmatrix} n & s \\ m & r \end{pmatrix}\xi\begin{pmatrix} s & j \\ r & i \end{pmatrix}
\end{align}
called the in-interaction scattering. In the appendix, we briefly detail how to compute such quantities.

\section{Coboson superposition ansatz for a two-pair system} \label{sec:apply}

We employ the formalism to tackle a balanced system of $2\uparrow + 2\downarrow$ fermions. The composite boson formalism, through the identity (\ref{cob_identity}), allows us to write the wave function of the system as a superposition of two-coboson states constructed from the one-coboson eigenstates, that is
\begin{align}\label{TwoPair_superposition}
    \ket{\psi^{(2)}} = \sum\limits_{i,j} C_{ij} B_{i}^\dagger B_{j}^\dagger\ket{0}
\end{align}
with $\ket{\psi^{(2)}}$ being the wave function of the system.

\subsection{Single-pair eigenstates}

The building blocks of the formalism are states of fermion pairs. These states are given by the creation operators
\begin{align}
    B^\dagger_i &= \int dz_\alpha dz_\beta \textrm{ } \varPhi_i(z_\alpha, z_\beta) \hat{\Psi}^\dagger_{\uparrow}(z_\alpha)\hat{\Psi}^\dagger_{\downarrow}(z_\beta),
\end{align} 
where $\varPhi_i$ is satisfies $H\varPhi_i = E_i\varPhi_i$ with 
\begin{align}
    H = \sum\limits_{p = \alpha,\beta}\left( -\frac{1}{2} \frac{\partial^2}{\partial z_p^2} + \frac{1}{2} z_p^2 \right) + g\delta(z_\alpha-z_\beta).
\end{align}
Here, lengths and energies are expressed in natural units. Introducing center-of-mass/relative coordinates $z_{c/r} = \frac{1}{\sqrt{2}}(z_\alpha \pm z_\beta)$, the Hamiltonian separates and its eigenstates can be written as 
\begin{align}\label{cobosons_WF_1quantz}
    \varPhi_{n_c,\varepsilon}(z_\alpha,z_\beta) =
    \varphi_{n_c}(z_c)
    \phi^{(r)}_\varepsilon(z_r).
\end{align}
These states are labeled by quantum numbers $i = (n_c,\varepsilon)$ with 
\begin{align}
    E_i =E^{(c)}_{n_c} + E^{(r)}_{\varepsilon}.
\end{align}

The center of mass part is a simple harmonic oscillator $\varphi_{n_c}(z_c)$ with $n_c \in \mathbb{N}_0$ and $E^{(c)}_{n_c} = n_c + 1/2$. The relative part wave-function can be either an even or odd function, with $E^{(r)}_{\varepsilon} = \varepsilon + 1/2$. The odd solutions are given by $\phi^{(r)}_\varepsilon(z_r) = \varphi_\varepsilon(z_r)$ with $\varepsilon\in\mathbb{N}_{odd}$. The even solutions for the relative part have an energy $\varepsilon$, non-integer, and the wave functions are well known through the literature as parabolic cylinder functions \cite{avakian1987spectroscopy, franke2003coherence, busch1998two}. We, however, would like to determine a representation of these solutions with respect to a harmonic oscillator basis, i.e.
\begin{align}\label{RelativeWF_HOexpansion}
    \phi^{(r)}_\varepsilon(z_r) = \sum_{n_r=0}^{\mathcal{N}_r}A^{(\varepsilon)}_{n_r} \varphi_{n_r}(z_r),
\end{align}
where $\varepsilon$ and $A^{(\varepsilon)}_{n_r}$ are given by an exact diagonalization procedure depending on the truncation parameter $\mathcal{N}_r$. The relative motion energy spectrum is discrete and we use the index $\nu\in\mathbb{N}_0$ to label its energies, that is, $\varepsilon\equiv\varepsilon_\nu \in \left\{\varepsilon_0 < \varepsilon_1 < \varepsilon_2 < \dots \right\}$. This labeling respects the parity of the relative wave function $\phi^{(r)}_\varepsilon$, i.e. $\varepsilon_\nu = \nu$ for $\nu\in\mathbb{N}_{odd}$ and $\varepsilon_\nu$ is non-integer for $\nu\in\mathbb{N}_{even}$.

Rewriting the wave function (\ref{cobosons_WF_1quantz}) using the above expansion, one can then express the single-cosobon creation operator as 
\begin{align}\label{SingleCoboson_HOexpansion}
    B^\dagger_{n_c,\varepsilon_\nu} = \sum\limits_{n_\alpha,n_\beta}\braket{n_\alpha,n_\beta}{\varPhi_{n_c, \varepsilon_\nu}}d^\dagger_{n_\alpha n_\beta},
\end{align}
where the coefficients are given by
\begin{align}
    \braket{n_\alpha,n_\beta}{\varPhi_{n_c, \varepsilon_\nu}} = \int dz_\alpha dz_\beta \textrm{ } {\varphi_{n_\alpha}(z_\alpha)}^*{\varphi_{n_\beta}(z_\beta)}^*\varPhi_{n_c, \varepsilon_\nu}(z_\alpha, z_\beta).
\end{align}
The calculation of the above coefficients is greatly simplified by using the Talmi-Brody-Moshinsky formula \cite{robledo2010separable}. The number of harmonic oscillator states $n_\alpha, n_\beta$ included in (\ref{SingleCoboson_HOexpansion}) {is characterized by a cut-off value $n_b$ which depends on} the center-of-mass quantum number $n_c$ and the truncation parameter $\mathcal{N}_r$ in (\ref{RelativeWF_HOexpansion}). {Throughout the paper, $n_b$ varies according to the interaction regime, ranging from $n_b = 50$ in the strongly attractive limit to $n_b = 30$ for both weak and repulsive interactions.}

It will be useful to know how the single-pair eigenstates transform through the symmetries $\mathcal{P}$ and $U_{\mathcal{T}}$. From the parity of both center of mass and relative parts of (\ref{cobosons_WF_1quantz}), one sees that the single-pair creation operator transform as 
\begin{align}
    &\mathcal{P}B^\dagger_{n_c,\varepsilon_\nu}\mathcal{P}^{-1} = (-1)^{n_c+\nu}B^\dagger_{n_c,\varepsilon_\nu}\label{symm_P}\\
    &U_{\mathcal{T}}B^\dagger_{n_c,\varepsilon_\nu}{U_{\mathcal{T}}}^{-1} = (-1)^{\nu} B^\dagger_{n_c,\varepsilon_\nu}.\label{symm_U_T}
\end{align}
Hence, the single-pair eigenstates have inversion and spin-flip parity phases given by $P_{n_c,\varepsilon_\nu}  = (-1)^{n_c+\nu}$ and $T_{n_c,\varepsilon_\nu}  = (-1)^{\nu}$, respectively.

\subsection{Construction of the coboson superposition ansatz}

Having determined the single-coboson eigenstate basis, we construct the superposition (\ref{TwoPair_superposition}) by selecting which states two-cobosons states will be included based on considerations over the interaction regime and symmetries. 

The general form of the superposition is 
\begin{align}\label{Superposition_ansatz}
    \ket{\psi^{(2)}} = {\sum\limits_{\substack{n_c,\nu \\  \bar{n}_c,\bar{\nu}}}}^{\prime}C_{n_c \nu;\bar{n}_c\bar{\nu}}B^\dagger_{n_c, \varepsilon_{\nu}}B^\dagger_{\bar{n}_c\varepsilon_{\bar{\nu}}}\ket{0},
\end{align}
where the prime in the summation implies an appropriate truncation. An immediate constraint on the sums arises from the commutativity of coboson creation operators, realized by relation (\ref{commut_Kin_2}), which forces the coefficients $C_{n_c \nu;\bar{n}_c\bar{\nu}}$ to be symmetric under the exchange $(n_c, \nu) \leftrightarrow (\bar{n}_c,\bar{\nu})$.

Our construction of the ansatz is performed by a selection of states $B^\dagger_{n_c, \varepsilon_{\nu}}B^\dagger_{\bar{n}_c\varepsilon_{\bar{\nu}}}\ket{0}$ accordingly to the following constraints:
\begin{itemize}
    \item Relative quantum numbers $\varepsilon_{\nu},\varepsilon_{\bar{\nu}}$ are restricted to a small subset. 
    
    \item Center-of-mass quantum numbers satisfy $n_c + \bar{n}_c \leq \mathcal{N}_c$ with $\mathcal{N}_c$ being a truncation parameter. {This constraint is essentially an center-of mass energy truncation inspired by exact diagonalization methods with non-interaction energy cut-off \cite{plodzien2018numerically, pkecak2022unconventional, rammelmuller2023modular}}.
    
    \item Two-coboson states must have either $+1$ or $-1$ parity under $\mathcal{P}$ or $U_{\mathcal{T}}$. Using relations (\ref{symm_P}) and (\ref{symm_U_T}), one can see that the parity phases of the two-coboson states are $P =P_{n_c,\varepsilon_\nu}P_{\bar{n}_c\varepsilon_{\bar{\nu}}}$ and $T=T_{n_c,\varepsilon_\nu}T_{\bar{n}_c\varepsilon_{\bar{\nu}}}$.
\end{itemize}

{The number $\mathcal{N}_{sup}$ of two-coboson states in the superposition (\ref{Superposition_ansatz}) is given by the three constraints above.} In this work, we deal mainly with ground-state properties and the energy spectrum of some low-lying states.

\subsubsection*{Ground-state}

In order to study the ground state of the system, one symmetry constraint is that the wave function bear an even spatial inversion parity, that is $P = +1$. Further constraints will depend on the interaction regime.

For the attractive regime ($g<0$), we restrict the relative quantum numbers of the single-coboson states to its ground-state $\varepsilon_{0}$, thus yielding the ansatz
\begin{align}\label{CobAnsatz_GS_attrac}
    \ket{\psi^{(2)}_0} = \sum\limits_{\substack{n_c + \bar{n}_c \leq \mathcal{N}_c \\ P = +1 }}C_{n_c\bar{n}_c}B^\dagger_{n_c, \varepsilon_{0}}B^\dagger_{\bar{n}_c\varepsilon_{0}}\ket{0}.
\end{align}

For the repulsive regime ($g>0$), we restrict the relative quantum numbers of the single-coboson states to the subset $\{\varepsilon_{0}, \varepsilon_{2}, \varepsilon_{4}\}$, yielding 
\begin{align}\label{CobAnsatz_GS_repuls}
    \ket{\psi^{(2)}_0} = \sum\limits_{\substack{n_c + \bar{n}_c \leq \mathcal{N}_c \\ \nu,\bar{\nu}\in\{0,2,4\} \\ P = +1}}C_{n_c \nu;\bar{n}_c\bar{\nu}}B^\dagger_{n_c, \varepsilon_{\nu}}B^\dagger_{\bar{n}_c\varepsilon_{\bar{\nu}}}\ket{0}.
\end{align}

\subsubsection*{Low-lying states}

For determining the low-lying states in a weakly interacting regime, we restrict the relative quantum numbers of the single-coboson states to the subset $\{\varepsilon_{0}, \varepsilon_{1}, \varepsilon_{2}\}$ and select either the even ($T=+1$) or odd ($T=-1$) states with respect to the spin-flip symmetry. For such case, the ansatz bears the form
\begin{align}\label{CobAnsatz_LowLying}
    \ket{\psi^{(2)}} = \sum\limits_{\substack{n_c + \bar{n}_c \leq \mathcal{N}_c \\ \nu,\bar{\nu}\in\{0,1,2\} \\ T =\pm 1}}C_{n_c \nu;\bar{n}_c\bar{\nu}}B^\dagger_{n_c, \varepsilon_{\nu}}B^\dagger_{\bar{n}_c\varepsilon_{\bar{\nu}}}\ket{0}.
\end{align}

\subsection{Schrödinger equation}

The eigenstates of the system are obtained by solving the Schrödinger equation
\begin{equation}\label{exact_eigen_equation}
    \left( \hat{\mathcal{H}} - E^{(2)}\right)\ket{\psi^{(2)}} = 0.
\end{equation}
Substituting expansion (\ref{TwoPair_superposition}) in equation (\ref{exact_eigen_equation}) and applying $\bra{0}B_{q}B_{p}$ to the right-hand side, we obtain a generalized eigenvalue problem for the coefficients $C_{ij}$, namely
\begin{align}
	\sum\limits_{ij}
    \left[ 
    \mathcal{H}\begin{pmatrix} q & j\\ p & i\end{pmatrix} - E^{(2)}
    \mathcal{S}\begin{pmatrix} q & j\\ p & i\end{pmatrix}
    \right]C_{ij} = 0.
\end{align}
The following notations were introduced for the Hamiltonian and overlap matrices
\begin{align}
	&\mathcal{H}\begin{pmatrix} q & j\\ p & i\end{pmatrix} = 
    \bra{0}B_{q} B_{p}\hat{\mathcal{H}}B_{i}^\dagger B_{j}^\dagger\ket{0}\label{Ncoboson_Hamiltonian} \\
    &\mathcal{S}\begin{pmatrix} q & j\\ p & i\end{pmatrix} = 
    \bra{0}B_{q} B_{p}B_{i}^\dagger B_{j}^\dagger\ket{0}. \label{Ncoboson_OverlapS}
\end{align}

The above matrix elements can be calculated as follows. First, with relations (\ref{commut_Kin_1})-(\ref{commut_Kin_3}), we can calculate the overlap matrix, yielding
\begin{align}\label{OverlapS_explicit}
    \mathcal{S}\begin{pmatrix} q & j\\ p & i \end{pmatrix} 
    =\delta_{pi}\delta_{qj} - \PauliScatt{q}{j}{p}{i}+(i\leftrightarrow j).
\end{align}
Then, employing relations (\ref{H_B_commut})-(\ref{V_B_commut}), one can show that
\begin{align}
    \mathcal{H}\begin{pmatrix} q & j \\ p & i \end{pmatrix} &=\left(E_i + E_j\right)\mathcal{S}\begin{pmatrix} q & j \\ p & i \end{pmatrix} + \zeta\begin{pmatrix} q & j \\ p & i \end{pmatrix}
\end{align}
where $E_i$ gives the $i$-th pair eigenenergy and 
\begin{align}
   \zeta \begin{pmatrix} q & j \\ p & i \end{pmatrix} = \IntScatt{q}{j}{p}{i} - \xi^{in}\begin{pmatrix} q & j \\ p & i \end{pmatrix} + (i\leftrightarrow j),
\end{align}
with $\xi^{in}\begin{pmatrix} q & j \\ p & i \end{pmatrix}$ given by relation (\ref{in_interaction_Scatt}).

\section{Results} \label{sec:results}

In this section, we analyze key one-body and two-body observables, as density profiles and two-body correlators \cite{sowinski2021few} of the $2\uparrow + 2\downarrow$ system across a range of interaction strengths. Such observables are accessible in current few-body experiments by time-of-flight imaging \cite{bergschneider2018spin, bergschneider2019experimental, holten2021observation, holten2022observation, brandstetter2025emergent}. We derive expressions for these observables in terms of coboson quantities, which are detailed in the Appendix \ref{sec:appendix A}. Additionally, we compute the low-lying energy spectrum of the system for weak interactions and estimated the pairing gap. 

\subsection{Density profiles and two-body correlations}

One crucial observable to be addressed is the ground-state density profile of the system
\begin{align}\label{density_profile}
    n^{(1)}_{\sigma}(z) = \frac{1}{N_\sigma}\expval{\hat{\Psi}^\dagger_{\sigma}(z)\hat{\Psi}_{\sigma}(z)}
\end{align}
which can be interpreted as the probability density of finding a single fermion of species $\sigma$ sitting at position $z$. In the above expression, $N_\sigma$ is the number of fermions of species $\sigma$ and the expected value $\expval{\dots}$ is calculated with respect to the ground-state of the system. The above density profile is normalized to unity $\int_{-\infty}^{+\infty} n^{(1)}_{\sigma}(z)\textrm{ }dz = 1$.

There are two limiting cases where one can find analytic expressions for the density profiles \cite{grining2015crossover,grining2015many}. In the regime of strongly repulsive interaction, $g \rightarrow +\infty$, the density profile of the system behaves like that of a fully polarized gas of $N = 4$ fermions, analogous to a Tonks-Girardeau (TG) gas, 
\begin{align}\label{1body_dens_TG}
    n^{(1)}_{TG}(z) = \frac{1}{4}\sum\limits_{n=0}^{3}|\varphi_n(z)|^2, 
\end{align}  
where $\varphi_n(z)$ is the $n$-th harmonic oscillator function given by (\ref{HO_func_n}). In the regime of strongly attractive interaction, $g \rightarrow -\infty$, the density profile of the system behaves like that of a gas of $N = 2$ hard-core bosonic dimers analogous to a Lieb-Liniger (LL) gas,
\begin{align}\label{1body_dens_LL}
    n^{(1)}_{LL}(z) = \frac{1}{2}\sum\limits_{n=0}^{1}|\tilde{\varphi}_n(z)|^2, 
\end{align}  
where $\tilde{\varphi}_n(z) = 2^{\frac{1}{4}}\varphi_n(\sqrt{2}z)$ is the $n$-th harmonic oscillator function for a dimer of mass $2m$.

\begin{figure*}[hbt!]
    \includegraphics[scale = 0.32]{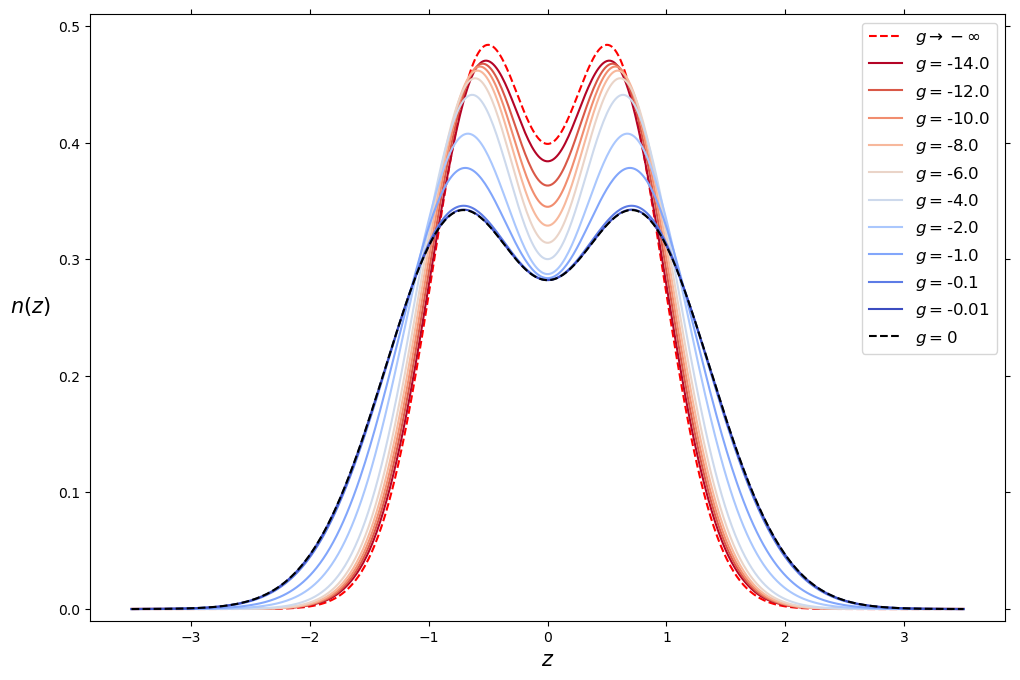}
    \includegraphics[scale = 0.32]{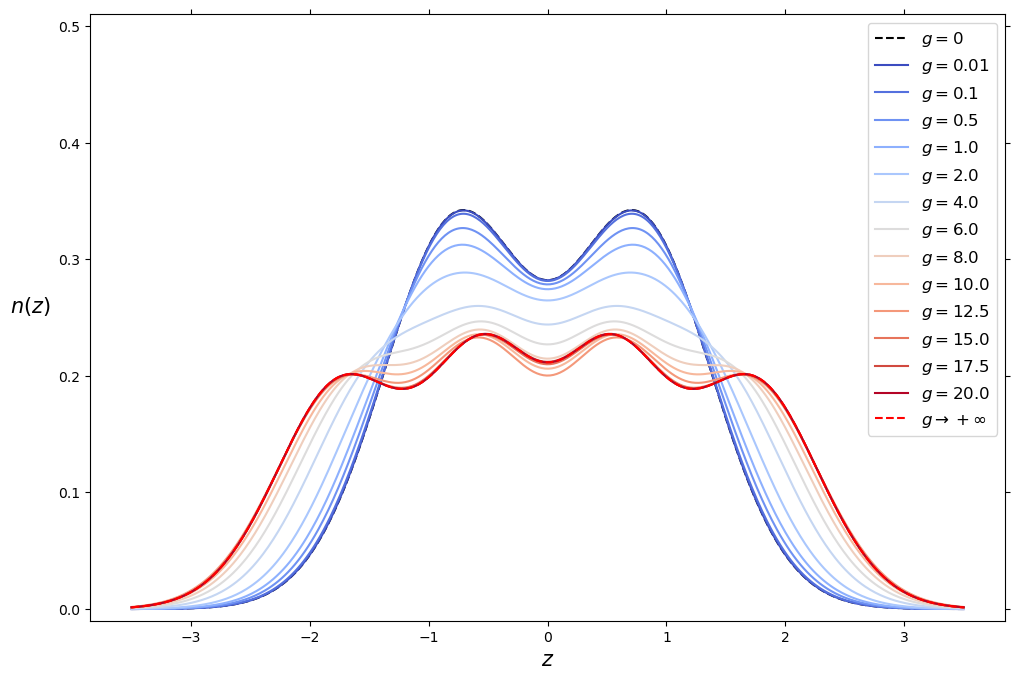}
    \caption{Ground state density profiles per spin species $n(z) = n_\uparrow(z) = n_\downarrow(z)$ for the $2\uparrow + 2\downarrow$ system. All quantities are expressed in natural harmonic oscillator units. 
    \textbf{Left}: Attractive regime with the dimensionless coupling $g$ taking values ranging from $g=0$ (non-interacting) to $g=-\infty$ (infinite attraction). The profile for the infinitely attractive regime is given by a Lieb-Lineger gas of $N = 2$ hardcore bosonic dimers.
    \textbf{Right:} Repulsive regime with the dimensionless coupling $g$ taking values ranging from $g=0$ (non-interacting) to $g=+\infty$ (infinite repulsion). The profile for the infinitely repulsive regime is given by a Tonks-Girardeau gas of $N = 4$ polarized fermions.}
    \label{graphs_onebody_dens}
\end{figure*}

In Fig. \ref{graphs_onebody_dens}, we plot the ground state density profiles per spin species $n(z) = n_\uparrow(z) = n_\downarrow(z)$ for interactions ranging from strong attraction to strong repulsion. The profiles for both species are identical since the considered system has equal number of both species. The density profiles obtained using the coboson superposition ansatz effectively capture the transition between the LL and TG regimes, offering a clear depiction of the system's behavior across varying interaction strengths. Furthermore, in the repulsive regime, our method captures the Fridel-Wigner transition characterized by a doubling of the peaks in the density profiles \cite{xianlong20122, kylanpaa2016thermal}.

For the attractive regime, we used the ansatz (\ref{CobAnsatz_GS_attrac}) with $\mathcal{N}_c = 25$ which yielded a superposition of $\mathcal{N}_{sup} = 91$ two-coboson states. For the repulsive regime, we employed an ansatz (\ref{CobAnsatz_GS_repuls}) with $\mathcal{N}_c = 8$ which yielded a superposition of $\mathcal{N}_{sup} = 120$ two-coboson states.

{Exact diagonalization methods employ wave functions constructed as a superposition of Fock states built from single-particle orbitals. These superpositions are truncated by limiting the number $n_b$ of single particle states or by imposing a non-interacting energy cut-off to the Fock states \cite{plodzien2018numerically}. For a $2\uparrow + 2\downarrow$ system, such method yields, with $n_b = 20$, a number of Fock states in the superposition ranging roughly from $\mathcal{N}_{sup} = 2000$ to $\mathcal{N}_{sup} = 30'000$ \cite{rammelmuller2023modular}.}

{We can compare the density profiles in Fig. \ref{graphs_onebody_dens} with profiles calculated with exact diagonalization methods, for example ref. \cite{grining2015many} (Figure 10a) and ref. \cite{rojo2020static} (Figure 5). Notice, however, that we plot the single-fermion density profiles normalized to unity and the mentioned papers plot the total density profiles normalized to  particle number. Our method yields comparable results with a relatively modest number of terms in the superposition (\ref{Superposition_ansatz}).}

For the case of Fermi gases, one also has interest in inter-component correlations. Such correlations can be described by diagonal part of the two-fermion reduced density matrix
\begin{align}\label{density_corr}
    n^{(2)}_{\uparrow\downarrow}(z_\alpha,z_\beta) = \frac{1}{N_\uparrow N_\downarrow} \expval{\hat{\Psi}^\dagger_{\uparrow}(z_\alpha)\hat{\Psi}^\dagger_{\downarrow}(z_\beta)\hat{\Psi}_{\downarrow}(z_\beta)\hat{\Psi}_{\uparrow}(z_\alpha)}.
\end{align}
The expectation value $\expval{\dots}$ is also calculated with respect to the ground-state. The above quantity can be interpreted as the probability of finding a spin-$\uparrow$ fermion at position $z_\alpha$ given a spin-$\downarrow$ fermion has been found at position $z_\beta$. This expression is also normalized to unity. 

Even for a non-interacting regime, distribution (\ref{density_corr}) still exhibits correlations due to the Pauli principle between the fermionic components of the gas. Thus, one can introduce the so-called noise correlation
\begin{align}\label{noise_corr}
    g^{(2)}_{\uparrow\downarrow}(z_\alpha,z_\beta) = n^{(2)}_{\uparrow\downarrow}(z_\alpha,z_\beta) - n^{(1)}_\uparrow(z_\alpha)n^{(1)}_\downarrow(z_\beta),
\end{align}
where the correlations due to the fermion indistinguishability is subtracted, leaving only correlations due to interactions \cite{sowinski2021few}.

\begin{figure*}[hbt!]
    \includegraphics[scale = 0.45]{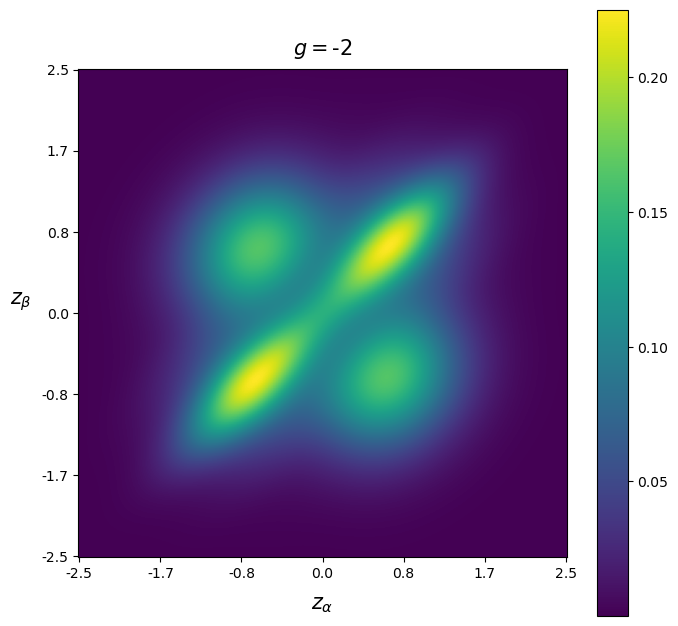}
    \includegraphics[scale = 0.45]{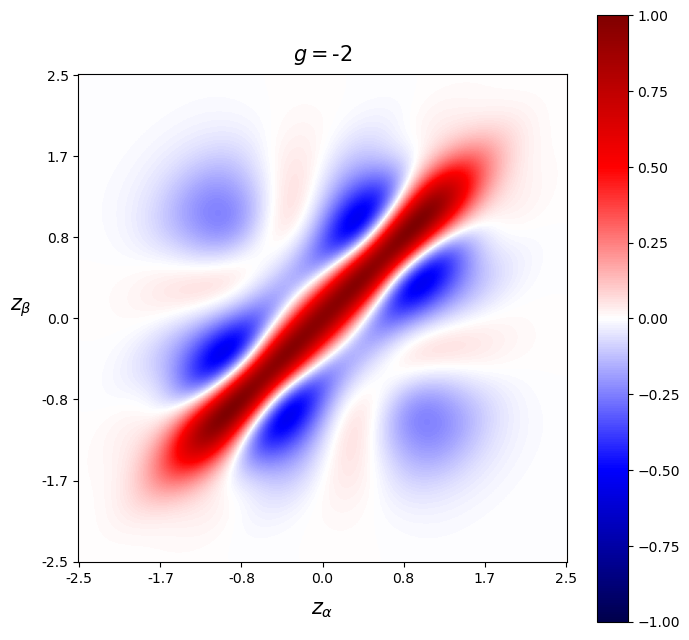}
    \caption{Ground state density correlations for the $2\uparrow + 2\downarrow$ system in the attractive regime with $g=-2$. Lengths are expressed in natural harmonic oscillator units. \textbf{Left}: Diagonal part of the two-fermion reduced density matrix $n^{(2)}_{\uparrow\downarrow}(z_\alpha,z_\beta)$. \textbf{Right}: Noise correlation $g^{(2)}_{\uparrow\downarrow}(z_\alpha,z_\beta)$. The color map for the noise correlation is expressed in arbitrary units and normalized such that the uncorrelated values correspond to 0 (white) as well as maximal positive correlation corresponds to 1.}
    \label{graphs_twobody_corr}
\end{figure*}

In Fig. \ref{graphs_twobody_corr}, we plot both density correlations for the ground-state in the attractive regime with $g=-2$. The ground-state was determined by a superposition of type (\ref{CobAnsatz_GS_attrac}) with $\mathcal{N}_c = 14$ which yielded a superposition containing $\mathcal{N}_{sup} = 36$ two-coboson states. The plot for $n^{(2)}_{\uparrow\downarrow}(z_\alpha,z_\beta)$ display a characteristic with respect to the $z_\alpha = z_\beta$ diagonal as well as two maxima along this line. We interpret this plot as depicting a higher probability of finding a $\uparrow\downarrow$-pair in symmetric positions with respect to the center of the trap. The plot for the noise correlation $g^{(2)}_{\uparrow\downarrow}(z_\alpha,z_\beta)$ is not so straightforwardly interpreted, however, we can compare it with plots obtained from other numerical methods \cite{rammelmuller2023magnetic}. 

\subsection{Low-lying spectrum and pairing gap}

\begin{figure*}[hbt!]
    \includegraphics[scale = 0.32]{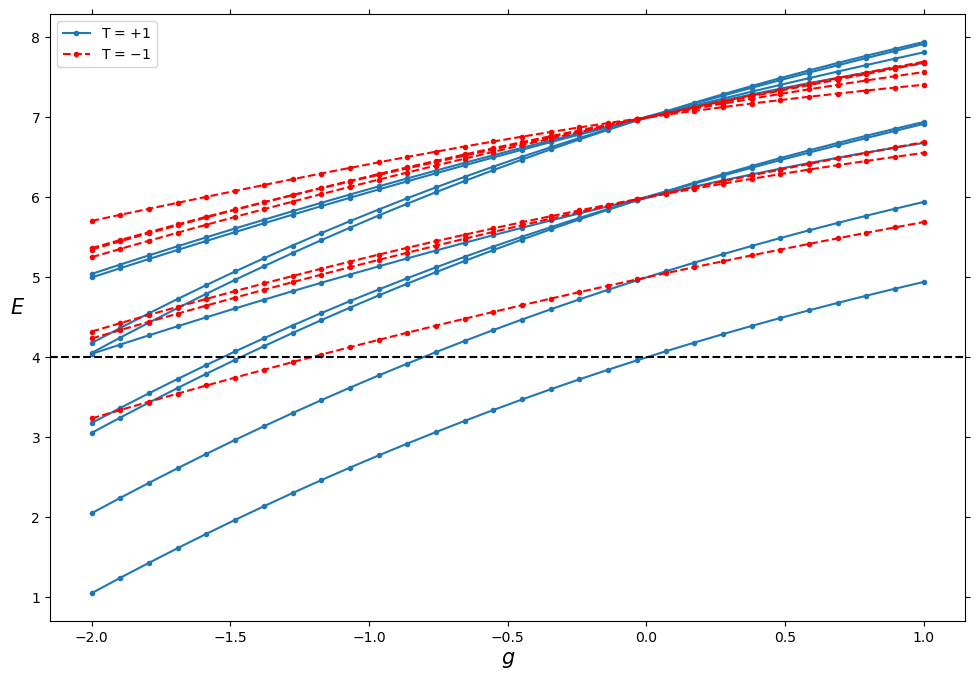}
    \includegraphics[scale = 0.32]{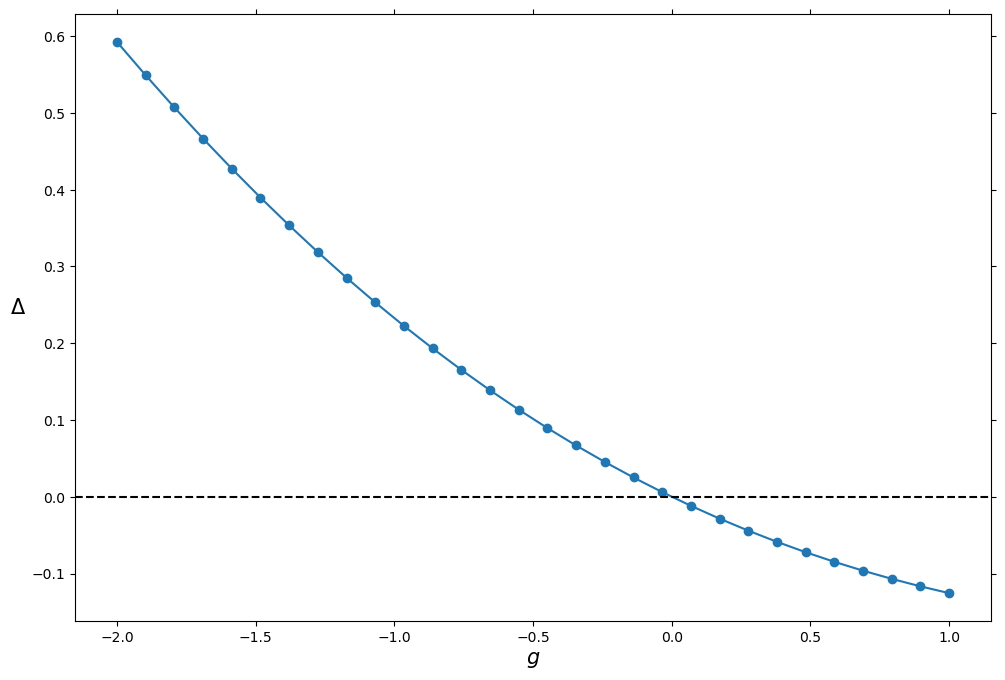}
    \caption{\textbf{Left}: Low-lying energy spectrum for the $2\uparrow + 2\downarrow$ system as a function of $g$. The blue solid curves show eigenstates with spin-flip parity $T=+1$ and the dashed red curves show states with spin-flip parity $T=-1$. The horizontal black dashed curve shows the level of the non-interacting energy. The difference between the lowest red dashed curve to the lowest blue solid curve is written as $2\Delta+1$, where $\Delta$ is the pairing gap. \textbf{Right}: Pairing gap for the $2\uparrow + 2\downarrow$ system as a function of $g$. The black dashed curve is added to help visualize when the pairing gap becomes negative, {which can be understood as a consequence of the finite size of the system}.}
    \label{graphs_lowEnergies_spinGap}
\end{figure*}

In the left panel of Fig. \ref{graphs_lowEnergies_spinGap}, we plot the low-lying energy spectrum of the system $2\uparrow + 2\downarrow$ as a function of the interaction strength $g$ resolved for both, even ($T=+1$) and odd ($T=-1$), symmetry sectors of the spin-flip symmetry $U_\mathcal{T}$. We used the ansatz (\ref{CobAnsatz_LowLying}) with $\mathcal{N}_c = 5$ which yielded a superposition of $\mathcal{N}_{sup} = 42$ states for the odd sector and of $\mathcal{N}_{sup} = 57$ states for the even sector. 

An important quantity that can be extracted from the energy spectrum is the pairing gap $\Delta$. Following Refs. \cite{d2015pairing, rammelmuller2023magnetic}, we determine $\Delta$ from the energy difference between the lowest red dashed curve (corresponding to the ground state in the $T = -1$ sector) and the lowest blue solid curve (the ground state in the $T = +1$ sector) in the left panel of Fig. \ref{graphs_lowEnergies_spinGap}, expressed as $2\Delta + 1$. The right panel of Fig. \ref{graphs_lowEnergies_spinGap} presents the dependence of the pairing gap $\Delta$ on the coupling constant $g$. {The negative values for $\Delta$ in the repulsive side can be understood as a consequence of the finite size of the system. The expected behavior at the thermodynamic limit is a vanishing pairing gap in the repulsive regime and a non-analytic transition at $g=0$ \cite{grining2015crossover}.}

\section{Concluding remarks} \label{sec:conclusion}

In this work, we investigated an one-dimensional harmonically trapped $2\uparrow + 2\downarrow$ system of half-spin fermions across a broad range of attractive and repulsive interactions. The coboson formalism provides an interesting framework for exploring many-body fermionic quantum systems while fully taking into account the fermionic nature of its components. From the solution of a single pair and symmetries of the system, we constructed a coboson superposition representation of the wave function of the many-pair system and computed key physical observables such as the particle density profiles, two-body correlations and the low-lying energy spectrum.

The calculated density profiles were consistent with well-known limits of the Tonks-Girardeau regime in the case of strong repulsion and the Lieb-Liniger regime under strong attraction, capturing the transition between these two regimes. Our method also captured the Fridel-Wigner transition of the density profile, in the strong repulsive regime, and provided sensible results for the two-body correlations and the low-lying energy spectrum. 

We advocate that this method is able to yield reasonable results offering an accessible alternative for the study of few-body systems and propose that one main advantage is the reduced number of states required in the superposition.

One way to further pursue this work could be to address larger fermionic few-body systems, such as \(3\uparrow + 3\downarrow\) and \(4\uparrow + 4\downarrow\), and explore dynamics beyond the ground state and correlations in low-dimensional quantum gases. It is also of interest to study spin imbalanced systems such as \(2\uparrow + 1\downarrow\) and \(3\uparrow + 2\downarrow\) \cite{patton2017trapped, pkecak2020signatures}, where a generalization of the formalism providing a coboson-fermion commutation algebra can be applied \cite{shiau2019coboson}. Studying these larger systems will allow us to investigate, for example, the behavior of the pairing energy and the even-odd effect \cite{sowinski2021few}. One can also explore systems with mass imbalance, that is, having component fermions of different masses \(m_\uparrow \neq m_\downarrow\) \cite{lydzba2020unconventional}. Collectively, these investigations can potentially provide further insight into quantum phase transitions and more complex pairing mechanisms \cite{rammelmuller2023magnetic}.

This study hints at the usefulness of the coboson superposition ansatz to model and analyze strongly correlated fermionic systems in reduced dimensions, thus contributing valuable perspectives to ongoing research in ultra-cold atomic physics and few-body quantum systems.

\section*{Acknowledgments} \label{sec:acknowledgements}

The authors would like to thank the Brazilian agency CAPES, for financial support. IR also thanks CNPq for partial support through contract 311876/2021.

\appendix
\section{Computation of key quantities and observables} \label{sec:appendix A}

In this appendix, we provide explicit formulas for calculating key coboson quantities and observables. Given a single particle basis, one can expand the single coboson eigenstates as 
\begin{align}\label{coboson_expansion}
    B^\dagger_i = \sum\limits_{n_\alpha,n_\beta}\braket{n_\alpha,n_\beta}{i} d^\dagger_{n_\alpha n_\beta}
\end{align}
with $d^\dagger_{n_\alpha n_\beta} := c^\dagger_{n_\alpha,\uparrow}c^\dagger_{n_\beta,\downarrow}$. Also, we introduce the following operators
\begin{align}
    & \mathcal{X}^{(\uparrow)}_{n \bar{n}} = \sum\limits_{n_\beta}\ket{n,n_\beta}\bra{\bar{n},n_\beta}\label{Chi_up}
    = \dyad{n}{\bar{n}} \otimes \mathds{1}_{\beta}\\
    &\mathcal{X}^{(\downarrow)}_{n \bar{n}} = \sum\limits_{n_\alpha}\ket{n_\alpha,n}\bra{n_\alpha,\bar{n}}\label{Chi_down} = \mathds{1}_\alpha \otimes \dyad{n}{\bar{n}}.
\end{align}
They act on the single-pair subspace and appear in the key formulas below.

\subsection{Computation of Pauli and Interaction Scatterings} 

In this section, we provide a brief exposition on the computation of the Pauli and interaction scatterings given by equations (\ref{PauliScatt_eq}),(\ref{interaction_Scatt}) and (\ref{in_interaction_Scatt}), respectively. 

The coboson eigenstate expansion (\ref{coboson_expansion}) allows us to write the Pauli scattering as 
\begin{align}\label{Pauli_Scatt_calc}
     \PauliScatt{f_2}{i_2}{f_1}{i_1} = \sum\limits_{n_\beta, \overline{n}_\beta}
     \bra{f_2}\mathcal{X}^{(\downarrow)}_{n_\beta \overline{n}_\beta}\ket{i_2}
     \bra{f_1}\mathcal{X}^{(\downarrow)}_{\overline{n}_\beta n_\beta}\ket{i_1}.
\end{align}
Using the above expression of the Pauli Scattering, we, then, can cast the interaction scattering into the form
\begin{align}
    &\xi\begin{pmatrix} f_2 & i_2 \\ f_1 & i_1 \end{pmatrix} = -\frac{1}{2}\left[
    \sum\limits_{n_\beta, \overline{n}_\beta}
     \bra{f_2}\mathcal{X}^{(\downarrow)}_{n_\beta \overline{n}_\beta}\ket{i_2}
     \bra{f_1}V\mathcal{X}^{(\downarrow)}_{\overline{n}_\beta n_\beta}\ket{i_1} + \right. \nonumber \\
    &+\left.\sum\limits_{n_\beta, \overline{n}_\beta}
     \bra{f_2}V\mathcal{X}^{(\downarrow)}_{n_\beta \overline{n}_\beta}\ket{i_2}
     \bra{f_1}\mathcal{X}^{(\downarrow)}_{\overline{n}_\beta n_\beta}\ket{i_1} + (i_1\leftrightarrow i_2)
    \right ],
\end{align}
where $V$ is the interaction potential, which acts in the single-pair subspace. From the above formulas, one can also compute the in-interaction scattering (\ref{in_interaction_Scatt}).

The "atomic" components of both expressions are brackets of the form $\bra{f}\mathcal{X}^{(\downarrow)}_{n_\beta \overline{n}_\beta}\ket{i}$ and $\bra{f}V\mathcal{X}^{(\downarrow)}_{n_\beta \overline{n}_\beta}\ket{i}$ calculated with respect to the known single-coboson eigenstates. Such brackets can be interpreted as matrices with respect to the indices $(n_\beta ,\overline{n}_\beta)$. If the expansion (\ref{coboson_expansion}) is built from $n_b$ single-fermion orbitals, the brackets will be $n_b\times n_b$ matrices. The expression (\ref{Pauli_Scatt_calc}) for the Pauli scattering, for example, can be implemented as 
\begin{align}
    \PauliScatt{f_2}{i_2}{f_1}{i_1} = \mathrm{Tr}\left\{ M_2 M_1 \right\}
\end{align}
with $M_k = \bra{f_k}\mathcal{X}^{(\downarrow)}_{\bullet\bullet}\ket{i_k}$. The other are analogous.

\subsection{Computation of reduced density matrices} 

In the current section, we provide formulas for calculating the first and second order reduced density matrices, $\rho^{(\sigma)}_{n\overline{n}}$ and $\rho^{(\uparrow\downarrow)}_{n_\alpha n_\beta;\overline{n}_\alpha\overline{n}_\beta}$, for the $2\uparrow + 2\downarrow$ system using the coboson formalism. These expressions are used to compute the density profiles (\ref{density_profile}) and density correlations (\ref{density_corr}).
Employing the superposition ansatz (\ref{TwoPair_superposition}), we can expand the reduced density matrices as
\begin{align}
    &\rho^{(\sigma)}_{n\overline{n}} = 
    \sum\limits_{\substack{f_1,f_2 \\ i_1, i_2 }}C_{f_1 f_2}^*C_{i_1 i_2}\expval{B_{f_2}B_{f_1}c^\dagger_{n,\sigma}c_{\overline{n},\sigma}B_{i_1}^\dagger B_{i_2}^\dagger}_0 \\
    &\rho^{(\uparrow\downarrow)}_{n_\alpha n_\beta;\overline{n}_\alpha\overline{n}_\beta} = 
    \sum\limits_{\substack{f_1,f_2 \\ i_1, i_2 }}C_{f_1 f_2}^*C_{i_1 i_2}\expval{B_{f_2}B_{f_1}d^\dagger_{n_\alpha n_\beta} d_{\overline{n}_\alpha\overline{n}_\beta}B_{i_1}^\dagger B_{i_2}^\dagger}_0, 
\end{align}
where the indices $i_k$ and $f_k$ label the single coboson states. Thus, one needs to compute the vacuum expectation values $\expval{B_{f_2}B_{f_1}c^\dagger_{n,\sigma}c_{\overline{n},\sigma}B_{i_1}^\dagger B_{i_2}^\dagger}_0$ and $\expval{B_{f_2}B_{f_1}d^\dagger_{n_\alpha n_\beta} d_{\overline{n}_\alpha\overline{n}_\beta}B_{i_1}^\dagger B_{i_2}^\dagger}_0$ which can be interpreted as matrix elements of the reduced density operators with respect to the two-coboson subspace.

To compute the expectation values mentioned, consider the following commutation relation:
\begin{align}\label{commut_n_B}
    &\left[c^\dagger_{n,\sigma}c_{\bar{n},\sigma}, B^\dagger_i\right] =\sum\limits_{j}\bra{j}\mathcal{X}^{(\sigma)}_{n\bar{n}}\ket{i}B^\dagger_j, 
\end{align}
where $\mathcal{X}^{(\sigma)}_{n\bar{n}}$ are given by (\ref{Chi_up}) and (\ref{Chi_down}).

Employing the above relations, it can be shown that 
\begin{align}
    \expval{B_{f_2}B_{f_1}c^\dagger_{n,\sigma}c_{\overline{n},\sigma} B_{i_1}^\dagger B_{i_2}^\dagger}_0 &= 
    \sum\limits_{l_1} \mathcal{S}\begin{pmatrix}f_2 & i_2\\ f_1 & l_1\end{pmatrix} \bra{l_1}\mathcal{X}^{(\sigma)}_{n \bar{n}}\ket{i_1}\nonumber\\
    &+ \sum\limits_{l_2} \mathcal{S}\begin{pmatrix}f_2 & l_2\\ f_1 & i_1\end{pmatrix} \bra{l_2}\mathcal{X}^{(\sigma)}_{n \bar{n}}\ket{i_2},
\end{align}
with $\mathcal{S}$ being a two-coboson state overlap given by (\ref{OverlapS_explicit}) and
\begin{align}
    &\expval{B_{f_2}B_{f_1} d^\dagger_{n_\alpha n_\beta} d_{\overline{n}_\alpha\overline{n}_\beta}B_{i_1}^\dagger B_{i_2}^\dagger}_0 = \nonumber \\
    &=\sum\limits_{k_1,l_1} \bra{f_1}\mathcal{X}^{(\downarrow)}_{n_\beta \overline{n}_\beta}\ket{k_1} 
                           \mathcal{S}\begin{pmatrix}f_2 & i_2\\ k_1 & l_1\end{pmatrix}
                           \bra{l_1}\mathcal{X}^{(\uparrow)}_{n_\alpha \overline{n}_\alpha}\ket{i_1} \nonumber \\
    &+\sum\limits_{k_1,l_2} \bra{f_1}\mathcal{X}^{(\downarrow)}_{n_\beta \overline{n}_\beta}\ket{k_1} 
                           \mathcal{S}\begin{pmatrix}f_2 & l_2\\ k_1 & i_1\end{pmatrix}
                           \bra{l_2}\mathcal{X}^{(\uparrow)}_{n_\alpha \overline{n}_\alpha}\ket{i_2} \nonumber \\
    &+\sum\limits_{k_2,l_1} \bra{f_2}\mathcal{X}^{(\downarrow)}_{n_\beta \overline{n}_\beta}\ket{k_2} 
                           \mathcal{S}\begin{pmatrix}k_2 & i_2\\ f_1 & l_1\end{pmatrix}
                           \bra{l_1}\mathcal{X}^{(\uparrow)}_{n_\alpha \overline{n}_\alpha}\ket{i_1} \nonumber \\
    &+\sum\limits_{k_2,l_2} \bra{f_2}\mathcal{X}^{(\downarrow)}_{n_\beta \overline{n}_\beta}\ket{k_2} 
                           \mathcal{S}\begin{pmatrix}k_2 & l_2\\ f_1 & i_1\end{pmatrix}
                           \bra{l_2}\mathcal{X}^{(\uparrow)}_{n_\alpha \overline{n}_\alpha}\ket{i_2}.
\end{align}

\clearpage
\bibliographystyle{apsrev4-1}
\bibliography{refs.bib}

\end{document}